\documentstyle[revtex]{aps}
\begin{document}
\begin{title}
On the Damping Rate of a Fast Fermion in Hot QED
 \end{title}
\author{R. Baier}
\begin{instit}
Institute for Theoretical Physics\\
University of California\\
Santa Barbara, CA 93106-4030, USA\\
\centerline{\it and}\\
Fakult\"at f\"ur Physik\\
Universit\"at Bielefeld\\
D-33501 Bielefeld, Germany
\end{instit}
\author{R. Kobes${}^*$}
\begin{instit}
Institute for Theoretical Physics\\
University of California\\
Santa Barbara, CA 93106-4030, USA\\
\centerline{\it and}\\
Lab.~de Phys.~Th\'eor. ENSLAPP\\
BP 110\\
F--74941 Annecy--le--Vieux\\
France
\end{instit}
\date
\maketitle
\begin{abstract}
The self-consistent determination of the damping rate of a fast
moving fermion in a hot QED plasma is reexamined. We argue how a detailed
investigation of the analytic properties of the retarded fermion Green's
function motivated by the cutting rules at finite temperature may resolve
ambiguities related to the proper definition of the mass-shell condition.
\end{abstract}
\pacs{PACS Numbers: 11.15.Bt, 12.38.Mh}
\section{Introduction}
Recent studies of damping rates of fast moving particles in a hot QED (or QCD)
plasma have raised some interesting problems [1--11]. The difficulties
start from the infrared sensitive behaviour of these rates: a logarithmic
divergence remains even after including the perturbative Braaten-Pisarski
resummation \cite{r12,r13}, which only screens the infrared sensitivity down to
scales of $O (eT)$. In the case of QED, which has no magnetic mass to
serve as infrared cut-off at the scale $O(e^2T)$, a self-consistent
determination of the rate $\gamma$ of a fast moving fermion in the plasma is
the
most elegant solution, as first suggested by Lebedev
and Smilga \cite{r2}.  Here $\gamma$ also plays the role of the infrared
cut-off.
In the analysis considered until now it is always assumed (even explicitly)
that the retarded propagator of the fast fermion (more precisely its
analytically continued form) has a true complex pole in the lower energy half
plane at the position $E-i\gamma$, with $E$ the energy of the fermion.
However, the presence of this pole does not allow an infrared stable solution
for $\gamma$ when the on-shell condition at this complex pole is
required \cite{r6}: this is indeed the favorable
 and consequent condition in the
case of a true pole on the ``physical'' Riemann sheet!

Based on the analytic structure of the retarded Green's functions as deduced
from  the general properties of spectral functions we argue in the following
that the complex pole in question is actually not on the ``physical'' sheet,
and we construct an explicit and simple example which should represent the
physical (realistic) situation
 and which may help to clarify the problem
of self-consistency for $\gamma$ \cite{r13a}.
By this attempt we take the point of view that
the damping rate of a fast moving fermion in a QED heat bath is a  physical
quantity, contrary to the arguments given in Ref.\cite{r8}.
Therefore an infrared finite result is to be aimed for it.

\section{Retarded Green's Function at Finite Temperature}
First it is convenient to illustrate some properties of the retarded Green's
function $G^R(p_0,p)$ [16--20].
It is defined as the boundary value of a complex function $G(z,p)$ as the
complex argument $z$ approaches the real axis from above:
\begin{equation}
G^R(p_0,p)\equiv G(z=p_0 +i\epsilon,p), \label{1}
\end{equation}
where
the function $G(z,p)$ is determined by a spectral function $\rho(p_0,p)$ as
\begin{equation}
G(z,p)= \int^{+\infty}_{-\infty} {dp'_0\over{p'_0 - z}} \rho(p'_0,p) .
\label{2}
\end{equation}
If $\rho(p_0,p)$ is a Lorentzian of the form
\begin{equation}
\rho(p_0,p)={\gamma/\pi\over(p_0-E(p))^2+\gamma^2} ,~~~ \gamma>0\,, \label{3}
\end{equation}
for {\it all\/} values of $p_0$, $-\infty<p_0<+\infty$, then the retarded
Green's function becomes
\begin{eqnarray}
G^R(p_0,p)& =&\int^\infty_{-\infty} {dp'_0\over p'_0-(p_0+i\epsilon)} ~~
{\gamma/\pi\over(p_0'-E(p))^2+\gamma^2}=\nonumber\\
& =&-{1\over p_0+i\epsilon-(E(p)-i\gamma)}, \label{4}
\end{eqnarray}
where we first keep $p_0$ on the real axis. However, this explicit function may
be analytically continued to complex values
 of $p_0$: for $Im\ p_0>0$~~$G^R$ is
regular as it should be from the definition (1), but for $Im\ p_0<0$ the
analytically continued $G^R$ has a true pole at $p_0=E-i\gamma$.
 The origin of this
pole is easily understood when considering the situation in the $p_0'$-plane
\cite{r20,r21}. As illustrated in Fig.~1a for the continuous integration
contour $-\infty<p'_0<+\infty$, the singularities of the integrand are at
$p_0'=p_0+i\epsilon$ and due to the Lorentzian at $p_0'=E\pm i\gamma$,
 indicated by
crosses. The contour for the retarded function is below the pole at
$p_0'=p_0+i\epsilon$, indicated by $\epsilon>0$. In order to continue
 $G^R$ into
the lower half plane, $Im\ p_0<0$, the contour
 has to be deformed (Fig.~1b), and
when $p_0'=p_0+i\epsilon$ approaches $p_0'=E-i\gamma$ it is pinched,
resulting in the
pole of $G^R$ at $p_0=E-i\gamma$. For the continuous contour (Fig.~1)
 this pinch is always present when the continuation to values for
$Im\ p_0<0$ is performed.

However, the situation becomes different when, for instance,
it is assumed that the spectral function $\rho$
is non-vanishing at some threshold $p_0\geq p_{th}$, but otherwise it is
zero. If, for example, $\rho$ is given for
$p_0\geq p_{th}$ by Eq.~(\ref{3}) with
$E(p)>p_{th}$, the retarded function is
\begin{eqnarray}
G^R(p_0,p)&=&\int^\infty_{p_{th}} {dp'_0\over p'_0-(p_0+i\epsilon)} ~~
{\gamma/\pi\over(p_0'-E(p))^2+\gamma^2}=\nonumber\\
 & =& {1\over2\pi i} \Bigg\{ {1\over p_0-(E(p)-i\gamma)} \ln
               \left[{ p_{th}-p_0-i\epsilon\over p_{th}-E(p)+i\gamma }
\right] \nonumber\\& &
 \qquad-{ 1\over p_0-(E(p)+i\gamma)}\ln \left[
{ p_{th}-p_0-i\epsilon\over p_{th}-E(p)-i\gamma
}\right] \Bigg\}. \label{5}
\end{eqnarray}
In the $p_0'$-plane the situation looks as in Fig.~2. One immediately sees
that when $p_0'=p_0+i\epsilon$ approaches $p'_0=E-i\gamma$, for example, the
$p_0'$-integration
contour is not necessarily pinched, and so there is {\it no\/} pole at
$p_0=E-i\gamma$ in the analytically continued $G^R$
on the first (``physical") sheet. Instead,
there is an endpoint singularity at $p_0=p_{th}$, which is a
logarithmic branch point, and the discontinuity across this cut on the
real axis is given (by construction) by the Lorentzian $\rho(p_0,p)$
of Eq.~(\ref{3}). With respect to this cut the pole has been moved
onto the other sheets obtained after continuation of the logarithm,
$\ln(p_{th}-p_0-i\epsilon)$: this amounts to deforming
 the contour in such a way that
a pinch is present (cf. Fig.~1b).

 The behaviour with respect to $p_0=E+i\gamma$ is symmetric
with respect to the one at $p_0=E-i\gamma$; on the first
sheet $G^R(p_0,p)$ of Eq.~(\ref{5}) is regular for
$Im\ p_0>0$ and for $Im\ p_0<0$, with the only singularity on this sheet
that of the logarithmic branch point at $p_0= p_{th}$.

The case just described is in close analogy to the discussion of resonances (at
zero temperature): with respect to the variable of
 the energy squared resonances
are on the unphysical sheet, and the branch point
 is determined by the threshold
properties of scattering amplitudes.

\section{Fermion Damping Rate}
As the simplest case we consider the damping rate of a heavy fermion (``muon")
of mass $M$ in a hot QED plasma of (massless) electrons and photons. The
energy, the momentum and the velocity of the ``muon" are denoted by $E, \vec
p ~~(p=|\vec p|)$, and $v$, respectively, but the main interest is in the limit
$v\to 1$ \cite{r6,r9,r10}.

Since we are only concerned with the leading order behaviour $e\to0$
at high temperature $T$, for a fast muon $(E\gg M>T)$ a couple of
approximations which simplify the calculations and the discussion may be
applied [1--10]. With $-(G^R)^{-1}=p_0-\Sigma^R$ and Eq.~(\ref{4})
we relate in the usual way $\gamma$ to the imaginary
part of the muon self--energy by $\gamma(p_0,p)=
-{\rm Im}\,\Sigma^R(p_0,p)$, with $p_0$ real. The
explicit expressions are derived, for example,
 in detail in \cite{r10}, which we closely
follow concerning conventions
and notation, and we start off with Eq.~(17) of this reference.
To leading order $v\to1$ we have
\begin{equation}
\gamma(p_0,p)\simeq +e^2  T \int_{{\rm soft}} {d^3q\over(2\pi)^3}
\int^{+q}_{-q}
{dq_0\over q_0} \rho_t (q_0,q) {\rm Im}\,\hat G^R (p_0- q_0,\vec p-\vec q).
\label{7}
\end{equation}
The following remarks summarize the results for $\gamma$
obtained so far in the literature:
\begin{itemize}
\item[--] Eq.~(\ref{7}) is derived in the one-loop
approximation, in which the hard
energetic ``muon" emits/absorbs one soft photon of four momentum $q^\mu$.
As well, corrections of order $1/E$ are neglected.
\item[--] Only the dominant transverse photon contribution has to be taken into
account, which following the Braaten-Pisarski hard thermal loop resummation
method \cite{r12} is determined by its spectral density $\rho_t(q_0,q)$; its
explicit form is given in \cite{r22}, but for the following we only require the
(approximate) integral
\end{itemize}
\begin{equation}
\qquad\quad\int^{+q}_{-q} {dq_0\over q_0} \rho_t(q_0,q)\simeq 1/q^2 \label{8}
\end{equation}
\begin{itemize}
\item[  ] when $q\leq eT$ \cite{r6,r9,r10}.
\item[--] All the complications due to the spin of the fermion are suppressed
in
Eq.~(\ref{7}), in the sense that the heavy fermion propagator is
described by a (retarded) scalar function ${\hat G}^R$.
\item[--] Inserting for a bare fermion
$-(\hat{G}^R)^{-1}=
p_0+i\epsilon-E(p),\ E(p)=
\sqrt{{\vec p}^{\,2}+M^2}$,
 into Eq.~(\ref{7}), we have
\end{itemize}
\begin{equation}
\qquad\quad{\rm Im}\,\hat G^R(p_0-q_0,\vec p-\vec q)\Big|_{q_0\to0}\simeq
 \pi\delta(p_0-\sqrt{ (\vec
p-\vec q)^2+M^2}),\label{9}
\end{equation}
\begin{itemize}
\item[  ] and one immediately finds the infrared divergent result
\end{itemize}
\begin{equation}
\qquad\quad
\gamma(p_0\simeq E,p)\simeq {e^2\over4\pi} T\int^{eT} {dq\over q}, \label{10}
\end{equation}
\begin{itemize}
\item[  ] This is mainly due to soft photon exchange, and it is
the origin of the
problems with the hard fermion damping rate. It also shows that to leading
order only the infrared sensitive region $|q_0|\leq q\to0$ has to be
considered. It is worth noting
that the coefficient in Eq.~(\ref{10}) is gauge parameter independent
\cite{r3,r5}.
\item[--] In hot QED there is no (non-perturbative) magnetic mass
to provide a cutoff to the logarithmic infrared divergence in
Eq.~(\ref{10})
 \cite{r18}. In QCD the magnetic mass is expected to be on the
scale $m_{mag}\simeq g^2T$, with $g$ the
strong coupling constant: consequently using $m_{mag}$ and evaluating
$\gamma$ on
the real axis $(p_0\simeq E)$ a finite value may be
 -- and has been -- argued for
quarks in QCD \cite{r3,r4,r7}.
\item[--] From Eq.~(\ref{10}) $\gamma$ is ``anomalous", in that
its magnitude is on the scale $e^2T$ (neglecting $\ln e$ factors for the
moment). Therefore it has been first conjectured
 by Lebedev and Smilga \cite{r2} to
use $\gamma$ itself as a possible infrared cutoff. This opens the
possibility of a self-consistent calculation of $\gamma$
by replacing ${\hat G}^R$ in Eq.~(\ref{7}) by the simple Lorentzian
of Eq.~(\ref{4}) \cite{r2}.
In Ref.\cite{r2} it is
shown that only the fermion propagator should be modified,
 but not the photon one; also {\it no\/} vertex corrections are required.
\item[--] The self-consistency requirement for the damping rate requires,
however, clarification of the ``on-shell" condition used to evaluate
Eq.~(\ref{7}).
On the one hand, as used in Ref.\cite{r2}, one can keep in Eq.~(\ref{7})
$p_0$ on the real axis, in which case $p_0= E(p)$ is used. On
the other hand,
one can generalize Eq.~(\ref{7}) to complex $p_0$ by demanding that it
holds at the complex pole $p_0=E(p)-i\gamma$
under the narrow
width assumption $\gamma \ll E(p)$.
 It is crucial
to point out that under the assumption that $G^R$ is given by
Eq.~(\ref{4}) the
complex pole is on the first (and only) sheet in the energy plane; therefore
the second case of self-consistency could be
argued as the physical one, and is in any case a point at which
gauge invariance can formally be proven to hold \cite{r27}.
However, as remarked in Ref.\cite{r6}, in this case the
 infrared divergence is not screened by a
non-vanishing $\gamma$! Therefore, this attempt fails for QED when
Eq.~(\ref{4}) is
used as a model for the ``dissipative" retarded fermion propagator, and as
such the narrow width condition in the form
${\rm Im}\,\Sigma^R(p_0=E)\simeq {\rm Im}\,\Sigma^R(p_0=E-i\gamma)$ does not
hold.
\end{itemize}

Rather than concluding that the fermion damping in QED is an unphysical
quantity, and therefore not observable as argued in Ref.\cite{r8},
we take the point of view that the ansatz of Eq.~(\ref{4}) for the
retarded energetic fermion propagator $G^R$ does not reflect the proper
physical conditions. Instead it is realistic to use as a simple model
of the spectral function the form
given by Eq.~(\ref{5}), which allows for branch cuts on the real axis in
the energy plane.

\section{Spectral Function and Cutting Rules}
When discussing the location of branch points and cuts in self-energy
functions at nonvanishing temperatures it is useful to recall that Weldon
\cite{r23} has shown, starting from the one-loop approximation and
extrapolating to the many particle case, that the branch points are determined
by the $T=0$ masses of the particles in the heat bath.
One can see this by considering the spectral function directly,
where one notes that it is
evaluated -- indepen\-dent of $T=0$ or $T\neq0$
-- by using the energy eigenstates
of the (full) Hamiltonian, $H|n>=E_n|n>$. For example,
 for fermions the resulting
$\rho$ at finite $T$ has the following structure \cite{r16,r17}:
\begin{eqnarray} & &
\rho(p_0,\vec p~ ) \hat= \sum_{m,n} (2\pi)^4 \delta (p_0-(E_n-E_m))
 \delta (\vec p - (\vec p_n-
\vec p_m)) \nonumber\\& &
  \cdot ~ e^{-E_m/T} (1-e^{-p_0/T}) |\langle m|\psi|n\rangle|^2,
\label{a1} \end{eqnarray}
where $\psi$ denotes the Dirac field operator. Obviously the energies $E_n$ and
momenta $\vec p_n$ appearing in the $\delta$-functions are temperature
independent, and therefore so are
 the positions of the branch points according to the
cutting rules \cite{r23,r24}.

Although the positions of the cuts do not depend on temperature, the
discontinuities across the cuts become temperature dependent. Let us
consider the one--loop $g{\bar\psi}\psi\phi$ self--energy example
of Weldon \cite{r23}. The
cut structure for the fermion self--energy with mass
$M$ is reproduced in Fig.~3
in terms of the variable $s=p^2_0-{\vec p}^{\,2}$ -- as at zero temperature
a small (zero temperature) photon mass $\lambda$ is introduced, having in
mind the limit $\lambda\to0$ whenever allowed.
The cut starting at
$s\geq(M+\lambda)^2$ is familiar from zero temperature. The cut between
$-(M^2-\lambda^2)\leq s\leq (M-\lambda)^2$ is due to the absorption/emission
of photons
from the heat bath, and vanishes for $T\to0$; in the limit $M>T$ its
discontinuity is exponentially suppressed, and therefore this cut is neglected
for the following discussion.
In Fig.~4 we plot the discontinuity in this example
for both hard and soft regions of external momenta -- we consider
the two contributions $A(p_0,p)\gamma_0 p_0$ and $M D(p_0,p)$
to ${\rm Im}\, \Sigma^R(p_0,p)$. We have kept the photon mass
$\lambda$ small but finite in these figures in order to differentiate
between the two regions $s>(M+\lambda)^2$
and $s<(M-\lambda)^2$ -- this is indicated by the break in the curves.
In the soft regime of Fig.~4(a)
we find that the Landau damping contribution for $s<(M-\lambda)^2$
dominates, as expected, but in the hard regime of Fig.~4(b)
the cut coming from $s>(M+\lambda)^2$ starts to dominate.

 From this we deduce that the thermal spectral function for the heavy,
energetic fermion propagator $-(G^R)^{-1}=p_0-\Sigma^R(p_0,p)$
has to have a contribution from at least the branch cut for
$s  \geq(M+\lambda)^2$, i.e. for $|p_0|\geq
\sqrt{{\vec p}^{\,2}+(M+\lambda)^2}$.
Note that, in this region, such a contribution
comes entirely from the self--energy $\Sigma^R$, since
there is no pole contribution.
Consequently we take as a realistic ansatz for the determination of $\hat
G^R$ in Eq.~(\ref{7}) the following:
\begin{itemize}
\item[(i)] for positive energy there is a single cut starting at
  $p_{th}=\sqrt{{\vec p}^{\,2}+(M+\lambda)^2}$; this is smaller than the
energy of the thermally excited heavy fermion, which receives contributions of
$O(eT)$ \cite{r22,r25,r26}, such that $E(p)-p_{th}\simeq O(e^2 T^2/E)$
for $\lambda\to0$ and $E\geq T$. Except for this, in the following the
thermal contribution of $O(eT)$
to the mass is neglected because of the limit $M>T$;
\item[(ii)] because of the narrow width condition $\gamma\ll E(p)$,
near $p_0\simeq E(p)$ the cut's discontinuity is dominated by the
nearby ``pole". We assume the
pragmatic parametrization of this discontinuity is given by the Lorentzian
of Eq.~(\ref{3}),
assuming $\gamma$ to be momentum independent (as a calculational
simplification);
\item[(iii)] in order to respect the symmetry properties of the fermion
spectral
density \cite{r24b}, $\rho(-p_0,\vec
p)=-\rho(p_0,\vec p)$, the following ansatz is suggested:
\end{itemize}
\begin{equation}
\qquad\quad
\rho(p_0,\vec p)={\gamma\over2\pi E(p)} \left[{1\over(p_0-E(p))^2+\gamma^2} -
  {1\over(p_0+E(p))^2+\gamma^2}\right] \Theta(|p_0|-p_{th}). \label{a2}
\end{equation}
\begin{itemize}
\item[  ] The temperature dependence only shows up in $\gamma=\gamma(T)$;
\item[(iv)] under the strong assumption -- which we accept for simplicity in
the
following -- that Eq.~(\ref{a2}) dominates $\rho$ for
 all values of $p_0$, and not only
in the neighbourhood of $E(p)$, one may require the sum rule to be
satisfied by the ansatz (\ref{a2}):
\end{itemize}
\begin{equation}
\qquad\quad1=\int^{+\infty}_{-\infty} \rho(p_0,\vec p)
\,p_0\,dp_0; \label{a3}
\end{equation}
\begin{itemize}
\item[  ] this fixes the normalization given in Eq.~(\ref{a2}) when terms of
$O(\gamma/E)$ are
neglected. This assumption leads to an overestimate of the
damping rate when determined self consistently.
\end{itemize}
Considering these points, we take as
the retarded Green's function for this simple toy model
that following from the spectral function of Eq.~(\ref{a2}) \cite{r24d}:
\begin{eqnarray} & &
G^R (p_0,\vec p)  = {1\over4\pi E(p)} \Bigg\{\Bigg[ {i\over
         p_0+i\epsilon-E(p)+i\gamma} \ln \left( {p_{th}+E(p)-i\gamma\over
        p_{th}-E(p)+i\gamma} ~~ {p_{th}-p_0-i\epsilon\over
	p_{th}+p_0+i\epsilon}\right) \nonumber\\& &
- (\gamma \to-\gamma)\Bigg]+ \Big[E(p)\to-E(p)\Big]\Bigg\}.
\label{a4}\end{eqnarray}
This function -- more precisely its analytic continuation in
$p_0$ -- does not have poles at $p_0=\pm E \pm i\gamma$ on the first
``physical" sheet with respect to the branch points at $p_0=\pm p_{th}$. The
discontinuity across the cut starting at $p_0=\pm p_{th}$ is given -- by
construction -- by the Lorentzian of Eq.~(\ref{a2})
 for real values of $p_0$, and near
$p_0\simeq E(p)$ we find, for $\gamma\ll E(p)$ and $E\ge T$,
 the narrow width condition
\begin{equation}
{{\rm Im}\,G^R(p_0=E(p))\over {\rm Im}\,G^R(p_0= E(p)-i\gamma)}
\simeq 1.
\label{a5}\end{equation}

\section{ Self-consistency Formulation}
Although we have assumed the spectral function of the propagator
 has the form of Eq.~(\ref{a2}) for
all values of $p_0$, in the true situation we might expect that this
would be only in a neighbourhood of $p_0\simeq E(p)$. Thus, for
a self--consistent determination of $\gamma$ we insert $G^R$ of
Eq.~(\ref{a4}) into Eq.~(\ref{7}) and evaluate the result
at the point $p_0=E(p)$; to leading order we then have
\begin{eqnarray} & &
\gamma(p_0\simeq E(p),\vec p)  \simeq {e^2\over4\pi^2} T \int^{eT}_0 dq
\int^{+1}_{-1} d\cos\theta {\gamma\over\gamma^2+q^2\cos\theta} \nonumber\\& &
\simeq {e^2\over2\pi^2} T \int^{eT}_0 {dq\over q}\arctan (q/\gamma) \simeq
{e^2\over4\pi} T\int^{eT}_\gamma {dq\over q} \nonumber\\& &
\simeq {e^2\over4\pi} T\ln {eT\over\gamma}\simeq {e^2\over4\pi} T\ln {1\over
e}\label{a6} \end{eqnarray}
where the infrared ``screening" by $\gamma$ is explicitly exhibited.
This then reproduces the original
self-consistent derivation of the fast damping rate in
Ref.\cite{r2}, but in this case without a singularity if one had
used instead the point
$p_0=E-i\gamma$ on the physical sheet to evaluate Eq.~(\ref{7}).
 This is consequently consistent
with the narrow width assumption of the form
${\rm Im}\,\Sigma^R(p_0=E(p))\simeq
{\rm Im}\,\Sigma^R(p_0=E(p)-i\gamma)$.

With minor modifications the preceding mechanism should be applicable to the
case of QCD fast damping rates and colour
relaxation times \cite{r28},
 without having to introduce a magnetic mass as an
infrared cut-off.

\section{Large Time Behaviour and Discussion}
In order to interpret $\gamma$ of Eq.~(\ref{a6}) in
the context of this toy model,
we study the time dependence of the Green function $G^R(t)$.
 This is given by the Fourier
transform of the retarded Green's function of Eq.~(\ref{a4}),
\begin{equation}
G^R(t)=\int^{+\infty}_{-\infty} dp_0 G^R (p_0)e^{-ip_0t}, \label{A1}
\end{equation}
where we now suppress the reference to the spatial momentum $\vec p$. Using the
spectral representation of $G^R(p_0)$ we obtain for
times $t$ real and positive
\begin{eqnarray}
G^R(t)&=& \int^{+\infty}_{-\infty}dp'_0 ~\rho (p'_0)\int^{+\infty}_{-\infty}
{e^{-ip_0t}\over-p_0+p'_0-i\epsilon}dp_0\nonumber\\
&=&2\pi i\int^{+\infty}_{-\infty} dp_0 ~\rho(p_0)e^{-ip_0t}.
\label{A2}   \end{eqnarray}
Inserting now the spectral function under consideration, Eq.~(\ref{a2}),
into Eq.~(\ref{A2}), we find
\begin{eqnarray}
G^R(t)&=& 2\pi i {\gamma\over2\pi E}\int^\infty_{p_{th}} dp_0 {e^{-ip_0t}
        \over(p_0-E(p))^2+\gamma^2} +c.c. \nonumber\\
&=&{2\over E}\left\{ \pi \sin(Et)e^{-\gamma t}
-\int_{(E-p_{th})/\gamma}^{(E+p_{th})/\gamma}
dx {\sin[(E-\gamma x)t] \over x^2+1}\right\} \nonumber\\
&\approx&{1\over E}\left\{ \pi \sin(Et)e^{-\gamma t}
+\cos(Et)\left[e^{-\gamma t} Ei(\gamma t)
-e^{\gamma t}Ei(-\gamma t)\right]\right\},
\label{A3} \end{eqnarray}
where $Ei(x)$ is the exponential integral \cite{r29} and the
approximation $E\gg T$, for
which $(E-p_{th})\ll \gamma$ and $(E+p_{th})\gg \gamma$, has been used.
One can consider Eq.~(\ref{A3}) in two limits -- if we assume
$\gamma t \gg 1$ then we find
\begin{equation}
G^R(t)\sim {2\over E} {\cos(Et)\over \gamma t} +
O\left({1\over \gamma t}\right)^2,
\end{equation}
while if we assume $\gamma t \ll 1$ then we obtain
\begin{equation}
G^R(t)\sim {\pi \over E} \sin(Et) +{\gamma t\over E}\left[
2(\ln(\gamma t)+1-\gamma_E)\cos(Et)-\pi\sin(Et)\right]+
O(\gamma t)^2,
\end{equation}
where $\gamma_E$ is Euler's constant.
Thus, as noted in
Ref.\cite{r16} and stressed in Ref.\cite{r8}, the time
 dependence of $G^R(t)$ may not
necessarily be of an exponential form, even for very large times,
and so care
must be taken in these cases in characterizing $\gamma$ as an
exponential ``damping''
rate. Even so, it is still a parameter within the context of the ansatz
for the Green function which remains to be determined, and for this we
can use the self--consistent condition derived from Eq.~(\ref{a2}):
$\gamma\sim -{\rm Im}\,\Sigma^R(p_0=E)$; the relation
${\rm Im}\,\Sigma^R(p_0=E) \simeq {\rm Im}\,\Sigma^R(p_0=E-i\gamma)$ assures us
to this
order that using the ``complex'' on--shell condition $p_0=E-i\gamma$
will lead to the same self--consistent determination of $\gamma$.
\acknowledgments
Fruitful discussions with P.~Aurenche,
 H.~Chu, A.~Ni\'egawa, S.~Peign\'e,
R.~D.~Pisarski, D.~Schiff, and A.~Smilga are gratefully acknowledged.
We also thank P.~Aurenche for a critical reading of the manuscript.

We thank with pleasure J.~Kapusta and E.~Shuryak for making possible a
stimulating participation at the ITP Workshop on ``Strong Interactions at
Finite Temperatures".

This research was supported in part by the National Science Foundation under
Grant No. PHY89-04035. Support by NATO under Grant CRG.930739, the
Natural Sciences and Engineering Research Council of Canada, and
the Centre International des Etudiants et Stagiaires de France
 is also acknowledged.

\vfill\eject

\vfill\eject

\figure{ Integration contour for the retarded Green's function.
Poles of the Lorentzian form of Eq.~(\ref{3}) are indicated by $\otimes$.
In (a) $p_0$ is real, while (b) shows the pole
singularity for $Im\ \ p_0<0$ which
 arises from the pinching of the contour between the two
singularities $p'_0=p_0$ and $p_0'=E-i\gamma$ of the integrand.}

\figure{Integration contour for Eq.~(\ref{5}), $p_{th}\leq p_0'<\infty$,
indicating that there is no pinch singularity on the first sheet at
$p_0'=p_0-E\mp i\gamma$ with respect to the endpoint singularity at
$p_0=p_{th}$.}

\figure{Location of the branch cuts (in the one-loop approximation) for the
thermal fermion propagator \cite{r23}. $\lambda$ denotes the $T=0$ photon mass
$(\lambda\to0)$, while $M$ is the fermion mass.}

\figure{The discontinuity of the one--loop fermion self--energy example
of Weldon \cite{r23}. In (a) we consider the soft region, with
$M/2T=0.2$ and $p/2T=0.1$, while in (b) we consider the hard region,
with $M/2T=15$ and $p/2T=10$.
The upper line in both figures
denotes $A(p_0)/A(p_0=\infty)$, while the lower line
denotes $D(p_0)/D(p_0=\infty)$.}

\clearpage
\pagestyle{empty}
\normalsize
\begin{center}
\par\noindent
\setlength{\unitlength}{0.240900pt}
\ifx\plotpoint\undefined\newsavebox{\plotpoint}\fi
\sbox{\plotpoint}{\rule[-0.175pt]{0.350pt}{0.350pt}}%

\par\noindent
\vspace*{11pt}
\par\noindent
$\qquad\qquad\qquad\qquad\qquad\qquad\qquad$  (a)
\vspace*{5pt}
\par\noindent
\setlength{\unitlength}{0.240900pt}
\ifx\plotpoint\undefined\newsavebox{\plotpoint}\fi
\sbox{\plotpoint}{\rule[-0.175pt]{0.350pt}{0.350pt}}%

\par\noindent
\vspace*{13pt}
$\qquad\qquad\qquad\qquad\qquad\qquad\qquad$  (b)
\vspace*{17pt}
\par\noindent
$\qquad\qquad\qquad\qquad\qquad\qquad\qquad$ Fig. 4
\end{center}


\begin{references}
\bibitem{r24b} The relevant density is the one related to the
Dirac operator, usually denoted by $\rho_1$ \cite{r21}.
\end{references}
\end{document}